\documentclass[eqsecnum,aps,prd,nofootinbib]{revtex4}

\usepackage{graphicx}%
\usepackage{enumerate}%
\usepackage{amsmath}%
\usepackage{amsfonts}
\def\be{\begin{equation}}
\def\bea{\begin{eqnarray}}
\def\ee{\end{equation}}
\def\eea{\end{eqnarray}}

\def\l: { \!:\!\!   }
\def\r: {  \!\!:\!  }
\def\lt: { :\!\!   }
\def\rt: {  \!\!:  }

\def\openone{\leavevmode\hbox{\small1\kern-3.3pt\normalsize1}}%
\def\simle{
    \mathrel{\rlap{\raise 0.511ex
        \hbox{$<$}}{\lower 0.511ex \hbox{$\sim$}}}}

\begin{document}

\title{Holographic Thought Experiments}

\author{Donald Marolf}

\affiliation{Physics Department, UCSB, Santa Barbara,  \\
CA 93106, USA \\\texttt{marolf@physics.ucsb.edu}}

\begin{abstract}The Hamiltonian of classical anti-de Sitter gravity is a pure boundary term on-shell.  If this remains true in non-perturbative quantum gravity then i) boundary observables will
evolve unitarily in time and ii) the algebra of boundary observables
is the same at all times. In particular, information available at
the boundary at any one time $t_1$ remains available at any other
time $t_2$. Since there is also a sense in which the equations of
motion propagate information into the bulk, these observations raise
what may appear to be potential paradoxes concerning simultaneous
(or spacelike separated) measurements of non-commuting observables,
one at the asymptotic boundary and one in the interior.  We argue
that such potentially paradoxical settings always involve a
breakdown of semi-classical gravity.  In particular, we present
evidence that making accurate holographic measurements over short
timescales radically alters the familiar notion of causality.  We
also describe certain less intrinsically paradoxical settings which
illustrate the above boundary unitarity and render the notion more
concrete.
\end{abstract}

\maketitle

\tableofcontents

\section{Introduction}

\label{intro}

Understanding quantum information in the context of black hole
evaporation is a long-standing issue in gravitational physics
\cite{HawkingRad}. One wishes to know whether information initially
sent into the black hole is again available after the evaporation is
complete and, if so, by what mechanism.  At least in the context of
string theory with anti-de Sitter (AdS) boundary conditions, the
advent of the AdS/Conformal Field Theory (CFT) correspondence
\cite{LargeN} appears to resolve at least the first question by establishing a
dual formulation in terms of a unitary field theory associated with
the AdS boundary.  In particular, this unitarity implies that the information can be recovered from operators in the dual theory and, by the usual rules assumed for AdS/CFT \cite{Witten}, such operators are associated with observables of the asymptotically AdS string theory at the AdS boundary.  Thus, in this context, it would appear that the information remains available after the evaporation is complete.

Nevertheless, an important puzzle remains: by what mechanism
and in what form does the information in the CFT remain available in the
gravitational description? Until this question is answered, some skepticism of the above-cited ``usual rules'' of AdS/CFT must necessarily remain.  Furthermore, there is a sense in which this AdS/CFT puzzle is even {\it more} acute than the original black hole question.  The intriguing point here is that AdS/CFT suggests that information sent into the spacetime through the AdS boundary at any early time $t_1$
remains available at the boundary at {\em any} later time $t_2 >
t_1$, whether or not enough time has passed for an energy flux
(Hawking radiation or otherwise) to return to the boundary; see figure 1.  It is this AdS puzzle that we will study below.

\begin{figure}
\includegraphics[width=5cm] {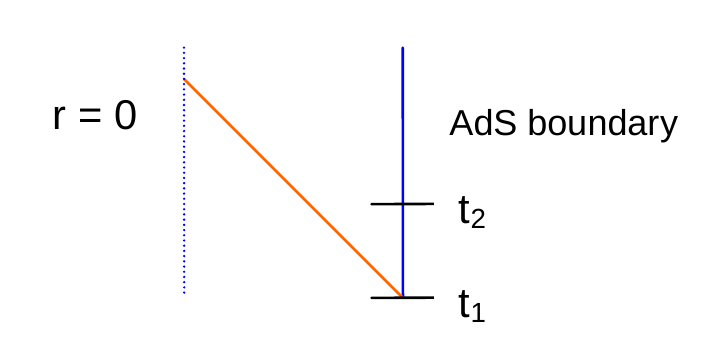}
 \caption{A conformal diagram of global AdS${}_{4}$ with the $S^2$ suppressed.  A signal leaves the boundary at time $t_1$.  The information is still present in the CFT at time $t_2$ though no signal has returned to the boundary.  }
 \label{signal}
 \end{figure}

A bulk explanation (reviewed in detail in section \ref{framework} below) of how the information can remain available at the boundary was recently offered in \cite{paper1}. Building on
\cite{BSP} and \cite{BMR}, it was noted that the desired properties follow
naturally if the on-shell quantum gravity Hamiltonian is a pure
boundary term.  In the classical theory, this well-known property follows directly from bulk diffeomorphism invariance.  The resolution of \cite{paper1} merely requires that the property continues to hold at the quantum level.
Now, many researchers expect that smooth spacetimes, and thus
diffeomorphism invariance per se, may play no fundamental role in the quantum theory.  However, there must be some structure that leads to diffeomorphism-invariance in the classical limit and whose consequences are similar.
It is plausible this quantum structure again implies that the on-shell Hamiltonian is a pure boundary term.

We shall follow \cite{paper1} in assuming that this is the case.  In
particular, we assume the Hamiltonian to be a self-adjoint generator
of time-translations on the boundary (though we make no a priori
commitment to the particular Hilbert space on which it is
self-adjoint).  By exponentiating this Hamiltonian, it follows immediately that the algebra of
boundary observables is independent of time and that information
present at an AdS boundary at any one time $t_1$ is also present
there at any other time $t_2$.  E.g., for systems invariant under time translations, any boundary observable ${\cal O}(t_1)$ at time $t_2$ can be represented as $e^{-iH(t_1-t_2)} {\cal O}(t_2)
e^{iH(t_1-t_2)}$ where ${\cal O}(t_2)$ is the same boundary observable at time $t_2$ and the Hamiltonian $H$ is also a boundary observable at time $t_2$.  An analogous statement holds in the time-dependent case; see appendix A.

This conclusion may cause some readers to question the extent to which the above assumptions are in fact reasonable.  Recall, however, that {\em without} making any assumptions,
\cite{paper1} also showed that perturbative gravity about a collapsing
black hole background is ``holographic'' in the sense that i) in the
asymptotically flat context a complete set of observables is
available within any neighborhood of spacelike infinity ($i^0$) and
ii) in the asymptotically AdS context, a complete set of observables
is contained in the algebra of boundary observables at each time
(technically, within any neighborhood of any Cauchy surface of the
conformal boundary).   The perhaps surprising conclusions to which our non-perturbative assumptions lead are thus established facts at the perturbative level, suggesting that these assumptions are worth investigating more deeply.

This is precisely the purpose of our work below.  We have three goals: to show more concretely the sense in which information is holographically encoded at the boundary, to begin to
investigate what sort of observers can access this information, and to
resolve certain potential paradoxes.  In particular, while
information remains present at the boundary as noted above, it is
clear that this information also propagates deep into the bulk. As
discussed in \cite{paper1}, there is no claim that quantum
information has been duplicated (which would violate the `no quantum
xerox theorem' \cite{noxerox}) but rather that the same qubit can be
accessed from two spacelike separated regions of spacetime.
Nevertheless, this raises interesting questions about non-commuting
measurements performed in the two regions: thinking of the qubit as
a single spin, what happens if an observer in the interior (say,
Bob) measures the $x$-component of the spin and a
spacelike-separated asymptotic observer (say, Alice) measures the
$z$-component? Similar issues were considered in \cite{comp,key,scramble}
with Bob inside a black hole, in which case it was argued that the
destruction of the interior observer at the black hole singularity
prevents comparison of these measurements and prohibits any true
contradiction. However, some other resolution is clearly required in
the absence of black holes, or more generally when Bob can
communicate with Alice.

The first class of measurements we study gives rise to just such potential paradoxes. Each
experiment involves a strong coupling to the Coulomb part of the
gravitational field, and in particular to a certain flux $\Phi$.  For reasons to be explained below, we refer to these experiments as the $\Phi$-subtraction protocol (section \ref{past}) and the $\Phi$-projection protocol (section \ref{projections}).  The couplings to $\Phi$ turn out to resolve the
apparent paradox by causing the usual semi-classical framework to
break down; such couplings are simply not compatible with smooth non-degenerate
metrics. Moreover, if such couplings can be described in some more
complete theory, we argue that this description would involve
a radical modification of the naive causal structure which allows Alice's measurement to affect Bob's results.  The second
class of experiments (section \ref{discrete}) is less intrinsically paradoxical,
but is consistent with smooth non-degenerate metrics.  As such, they
serve to make our notion of boundary unitarity more concrete.
Interestingly, these latter experiments rely on a certain
`operational density of states' being finite, while the measurements
of sections \ref{past} and \ref{projections} succeed without any such
assumption.  The general framework for our experiments is described in
section \ref{framework}, while the measurements themselves are analyzed in sections \ref{past}, \ref{projections}, and \ref{discrete}.  This part of our work will be based purely on bulk physics; no use will be made of AdS/CFT.  We then close with some final discussion in section
\ref{disc}.  In particular, section \ref{disc} {\it will} use AdS/CFT to suggest that, despite
taking us out of the usual semi-classical framework, the $\Phi$-projection protocol
of section \ref{projections} should nevertheless be allowed in a
full theory of quantum gravity.

Before beginning, we comment briefly on the issue of quantum fluctuations:  Our discussion above has assumed a definite causal structure for the space and ignored any quantum fluctuations of the causal structure.  This is in part because the issues of interest concern large weakly curved regions of spacetime near the AdS boundary where one would expect such quantum fluctuations to be small. Indeed, our main analysis below will make no explicit use of such quantum fluctuations. We therefore defer discussing the possible role of quantum causal structure fluctations until near the end of section \ref{disc}.

\section{A tale of two boundaries}
\label{framework}

The goal of this section is to set up a general framework useful for
discussing various holographic thought experiments.    Our main
concern will be diffeomorphism-invariance,  the gravitational
gauge-invariance.  This is clearly a key issue since, in the
classical theory, it is this symmetry that guarantees the
Hamiltonian to be a pure boundary term and leads to boundary unitarity.

As a result, we must be careful to measure only fully
gauge-invariant observables. The construction of
diffeomorphism-invariant observables is in general difficult in
non-perturbative gravity, but the task is greatly simplified by the
presence of a boundary. Typical boundary conditions (e.g., fixing
the boundary metric) break diffeomorphism-invariance so that the
behavior of bulk fields near the boundary readily defines
gauge-invariant observables.  This is true both at finite boundaries
and at asymptotic boundaries such as the AdS conformal boundary.  In
the second case, boundary operators are defined by suitably rescaled
limits of bulk fields as in e.g. \cite{Witten,Rehren}.  The reader
should consult these references for details; we will use this
construction without further comment.

We therefore place one observer (Alice) at, or perhaps more properly
outside, an asymptotic AdS boundary.  Aside from
Alice's measurements (discussed below), the boundary condition at
boundary A is of the familiar type which fixes the leading
Fefferman-Graham coefficient \cite{FG}.  E. g., in 3+1 dimensions we
take the metric near boundary A to be of the form
 \be
ds^2 = g_{ab} dx^a dx^b = \frac{\ell^2}{r^2}dr^2 + \left( g_{(0)CD} \frac{r^2}{\ell^2} + g_{(1) CD} \frac{r}{\ell} + g_{(2)CD} + g_{(3)CD} \frac{\ell}{r}
 +  \dots \right) dx^C dx^D,
 \ee
where $g_{(0)CD}$ is fixed and $g_{(1)CD}$, $g_{(2)CD}$ are
determined by $g_{(0)CD}$ and the Einstein equations. See e.g. \cite
{KSthermo} for various generalizations.  For simplicity, we consider
the case where $g_{(0)CD}$ takes the simple form
 \be
 \label{g0}
 g_{(0)CD} dx^C dx^D = - N_A^2 dt_A^2 + \Omega_{IJ} dy^I dy^J,
 \ee
with $y^I$ coordinates on $S^2$, $\Omega_{IJ}$ the round unit metric
on $S^2$, and $N_A$ a function only of $t_A$.  We will take $N_A$ to be a constant when Alice's couplings are turned off.

\begin{figure}
\includegraphics[width=8cm] {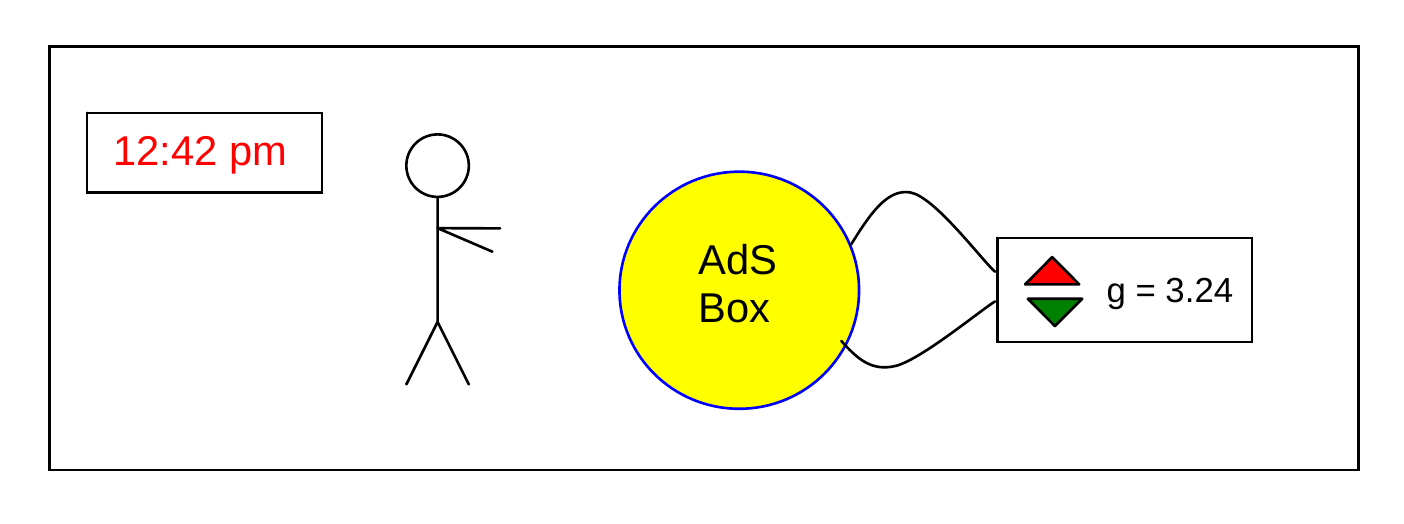}
 \caption{Our AdS system lives in a (conformal) box in Alice's laboratory.  Outside the AdS box are various ancilla.  A clock and a measuring device with adjustable coupling are shown.}
 \label{Lab}
 \end{figure}

We envision Alice as an
experimenter with the following characteristics:

\begin{enumerate}[i)]

\item{}  She has a notion of time evolution which coincides with that
of some preferred coordinate $t_A$ on the asymptotic boundary.
Reparametrizations of $t_A$ are not a gauge symmetry.

\item{} At her disposal are additional degrees of freedom
(ancilla) which are not part of the gravitating AdS spacetime.  We
encourage the reader to envision Alice as having a large
laboratory which {\em contains} the gravitating AdS system in a
(conformally compactified) box.  The ancilla are various useful
apparatus and quantum computers in this laboratory which exist outside the AdS
box.  See figure \ref{Lab}.

\item{}  Alice can couple her ancilla to AdS boundary observables
as described by any time-dependent
Hamiltonian.  Classically, this Hamiltonian is again a boundary term
(see appendix \ref{manyH} for details) and we assume this to be true
in the non-perturbative quantum theory as well.  A detailed example
of coupling the AdS space to such external degrees of freedom was
recently studied in \cite{Jorge}, though we will not need that level
of detail.

We will assume that Alice can choose the coupling arbitrarily, so long as it is local in $t_A$.  In
particular, we allow Alice to couple to boundary observables which are
non-local in space (e.g., integrals over $t_A = constant$
surfaces, spacelike Wilson lines, etc).  One might say that we impose
only a non-relativistic notion of causality on Alice's
ancilla\footnote{Some readers may desire a more concrete model which allows such couplings.  One
such model is to suppose that Alice's lab has more dimensions than
the AdS space, and that she can embed the AdS box in her lab in such
a way that events on the AdS boundary can be connected by causal
curves in her lab even when no such curve exists on the AdS boundary
itself.}.  We also allow such couplings to depend explicitly on $t_A$.  This gives Alice the ability to explicitly inject both information and energy into the AdS space (at, say, time $t_A=t_1$) which were not present in the AdS space before $t_A = t_1$.  A simple example is discussed in detail in appendix \ref{exB}.
\end{enumerate}
These assumptions provide an interesting and relatively simple
framework for exploration. We defer any discussion of the extent to
which they model a realistic observer to section \ref{disc}.

It remains to introduce our second experimenter (Bob). It might seem
natural to place Bob at Alice's boundary. However, doing so would
reduce any discussion of measurements to one familiar from
non-relativistic quantum mechanics.  The point is that, in this
case, Alice and Bob would share a common notion of time generated by
a common Hamiltonian $H$, and this Hamiltonian would transfer
information between the AdS space and both experimenter's ancilla.
The issues then boil down to the extent that we allow Alice and Bob
to couple to each other's ancilla.  For example, if Alice cannot
examine Bob's apparatus, then despite the unitarity of $e^{iHt}$ and
that fact that the information remains available to a sufficiently
boundary powerful observer, Alice simply does not have access to all
information and Bob's measurements will tend to disturb Alice's.
Similarly, Alice's measurements will tend to disturb Bob's.

On the other hand, placing Bob in the bulk raises two issues. First,
it becomes complicated to describe the gauge-invariant observables
to which Bob can couple. Second, such a placement raises the
possibility that all of Bob's apparatus may be holographically
encoded in boundary observables accessible to Alice. Alice then has
the ability to interact directly with Bob's ancilla, and in
particular to undo any measurement that Bob may have made.   In this
context no paradoxes need arise.

We therefore add a second (interior) boundary (B) to the AdS
spacetime.  We locate Bob at this boundary and endow him with
properties at boundary B in direct analogy with properties
(i,ii,iii) assumed for Alice at her boundary (A).  The one
difference between the two boundaries is that we take boundary B to
have a fixed {\em finite} metric. I.e., it is not an asymptotic
conformal boundary, but instead lies at a finite distance from
points in the interior.  This is a useful framework because
classical spacetimes allow signals respecting bulk causality to be
exchanged between the two boundaries.  In contrast, two asymptotic
AdS boundaries tend to be separated by horizons in any classical
solution, as occurs for example in the maximally extended AdS-Schwarzschild black
hole.  Such horizons limit (and plausibly remove) any settings for
potential paradoxes.

As we stress below and in appendix \ref{manyH}, even in the presence
of a second boundary the Hamiltonian boundary term at boundary A
generates time-translations along Alice's boundary alone.
Bob's boundary remains invariant. Similarly, the the Hamiltonian boundary
term at boundary B generates time-translations along Bob's boundary
but leaves boundary A invariant.  Again, these statements hold in
classical gravity and we assume they continue to hold at the
non-perturbative quantum level (in the same spirit as our original
assumption concerning the Hamiltonian as a boundary term). Readers
unfamiliar with these classical statements may see them most quickly by noting
that Gauss' law defines gravitational fluxes that are separately
conserved at each boundary when appropriate boundary conditions are imposed; further details and references are given in appendix \ref{manyH}.

As explained in detail below,
the result of the above assumptions is that information Alice injects into the AdS spacetime
through boundary A at time $t_1$ still remains available at boundary
A at time $t_2$ {\em no matter what Bob does at boundary B.} E.g.,
even if Alice injects the information as spins that travel to
boundary B where they are measured by Bob.  We investigate various such settings below.

We are most interested
in cases where Alice's measurement does not commute with Bob's.   In
sections \ref{past} and \ref{projections}, Alice performs a
holographic measurement at what appears to be a
spacelike separation from Bob's experiment, leading to the
potential paradox described in the introduction.  In particular, in
section \ref{past}, Alice attempts to directly measure the somewhat
artificial-looking observable $e^{-i\Phi_A(t_1-t_2)} {\cal O}(t_2)
e^{i\Phi_A(t_1-t_2)}$, where ${\cal O}(t_2)$ is a local boundary
observable at $t_A = t_2$ and $\Phi_A$ is the gravitational flux at
boundary $A$ which gives the associated boundary term in the Hamiltonian.  For reasons explained in section \ref{past}, we refer to this experiment as the $\Phi$-subtraction protocol.  Since, in the absence of Alice's
measurements, $\Phi_A$ is the full generator of $t_A$-translations,
this measurement allows Alice to recover information about ${\cal
O}$ an the earlier time $t_A = t_1$.  Despite the unfamiliar nature
of this experiement, it serves as a simple, clean example to
illustrate the consequences of Alice's coupling to $\Phi_A$:  Such
couplings necessarily alter the boundary conditions at boundary A
and, for large enough couplings of the right sign, are inconsistent
with smooth non-degenerate metrics. It is of course an open question
whether such couplings can be described in non-perturbative quantum
gravity and we save discussion of this issue for section \ref{disc}.
However, assuming for the moment that they are allowed, we argue in section
\ref{past} that they alter the naive notion of causality so that
Alice's measurement can in fact affect Bob's.

In section \ref{projections}, Alice performs a somewhat more
physical measurement, again at apparent spacelike separation from
that of Bob.  We refer to this experiment as the $\Phi$-projection protocol.  In rapid succession, Alice simply measures $\Phi_A$, a local
boundary observable ${\cal O}$, and $\Phi_A$ again, all with high
resolution. After a final interference experiment, and after
repeating this protocol many times on identically prepared AdS
systems, Alice obtains enough data to compute $A(E,\lambda,E') :=
\langle \Psi| P_{\Phi_A = E} P_{{\cal O}(t_2) = \lambda} P_{\Phi_A =
E'} |\Psi \rangle$. Here $|\Psi \rangle$ is the quantum state of the
system\footnote{\label{foot}Even if this state is not pure, there is
no harm is using notation appropriate to a pure state.  We may
consider the state to have been ``purified" by adding appropriate
ancilla.  Using pure state notation simplifies certain formulas in
section \ref{projections}.}, $P_{\Phi_A = E}, P_{\Phi_A = E'}$ are projections
onto the eigenspaces of $\Phi_A$ with eigenvalues $E,E'$, and
$P_{{\cal O} = \lambda}$ is the projection onto the eigenspace of
${\cal O}$ with eigenvalue $\lambda$. Integrating $A(E,\lambda,E')$
against $e^{-i(E-E')(t_1-t_2)}$, Alice computes $\langle \Psi |
P_{{\cal O}(t_1)} |\Psi \rangle$ and again recovers information
about ${\cal O}$ at any other time $t_1$.   However, the couplings
to $\Phi_A$ required for Alice to perform measurements of the
desired accuracy again impose boundary conditions inconsistent with
smooth invertible metrics and lead to the same discussion
as in section \ref{past}.

It is therefore of interest to ask if Alice can recover the
information using couplings compatible with smooth invertible bulk
metrics.  Section \ref{discrete} describes two experiments where
this is possible, provided that a certain `operational density of
states for Alice' is finite.  This density of states counts only
states distinguishable from boundary A, but allows Alice to reason
as if the spectrum of $\Phi_A$ were discrete.  The first
experiment is just a weak-coupling version of the $\Phi$-projection protocol in which Alice compensates for the weak coupling by letting
the experiment run for an exponentially long time.  Due to this
long time, her experiment is causally connected to Bob's, avoiding
the potential paradoxes of sections \ref{past} and
\ref{projections}.  In the second experiment,  Alice uses a generic
coupling to drain information from the AdS space into a universal
quantum computer (where she may then analyze the information at
will).  This experiment also requires enough time to make what is
effectively causal contact with Bob's measurement, though in
principal polynomial times will suffice.

\section{Measuring the past}
\label{past}

As described in section \ref{framework}, we consider two observers
(Alice and Bob), with Alice at an asymptotic (conformal) AdS
boundary (A) and Bob at a finite inner boundary (B).  We suppose both Alice and Bob to be interested in a qubit associated with the boundary value ${\cal O}$ of a local field at
time $t_1$; say, a spin degree of freedom, with ${\cal O}$ being the
$z$-component of the spin. The spin then travels inward and arrives at
boundary B. There Bob's apparatus detects the arrival of the spin
and measures some non-commuting observable (say, the $x$-component
$S_x$ of the spin), though it will not be necessary to model Bob's
measurements in detail.  For simplicity, it is perhaps best to consider a spin introduced at $t_1$ into the AdS space from outside.  In this case it is clear that Bob has no prior access to the spin.  As discussed in detail in appendix \ref{exB}, such an injection may be accomplished via a time-dependent coupling to one of Alice's ancilla.

As noted above, Alice wishes to couple directly to  $e^{-i\Phi_A(t_1-t_2)} {\cal O}(t_2)
e^{i\Phi_A(t_1-t_2)}$.  To model this measurement, it is convenient to write the AdS action in canonical form (see e.g. \cite{Waldbook}):
\begin{equation}
\label{Cact1} S_{total} = \int_{\Sigma \times {\mathbb R}}
\left(\pi \dot{\phi} - N {\cal H} - N^i {\cal H}_i \right) -
\int d t_A  N_A \Phi_A    + \int dt_B {\cal B}  ,
\end{equation}
where $\phi, \pi$ denote the full set of bulk fields and momenta,
including metric degrees of freedom, and a sum over fields is
implied.  We require no details of the B-boundary term ${\cal B}$ except that it is independent of both Alice's ancilla and the A-boundary observables.
We denote the usual lapse and shift by $N, N^i$ while ${\cal
H}, {\cal H}_i$ are the usual (densitized) bulk constraints, with
$i$ running over directions on a hypersurface $\Sigma$ of the AdS
space.  The boundary term $\Phi_A$ takes the usual form \cite{HT}
 \be
 \label{phi}
 \Phi_A = \frac{1}{16\pi G} \int_{S^2} d^2y \sqrt{\Omega}  \  \left( r^a P_{AdS}^{bc} D_b
 - r^b P_{AdS}^{ac} D_b  \right) g_{ac} ,
 \ee
where $r^a$ is a radial unit normal, $D_a$ is the covariant
derivative defined by a fixed metric $g^{AdS}_{ab}$ on exact
(global) anti-de Sitter space, and $P_{AdS}^{bc}$ is the projector
orthogonal to $\frac{\partial}{\partial t_A}$ defined by
$g^{AdS}_{ab}$.   This flux $\Phi_A$ can
also be written \cite{HIM} in terms of the boundary stress tensor of
\cite{HS,kraus} or in terms of the electric part of the Weyl tensor
at the A-boundary \cite{AM}.

We emphasize for later use that \eqref{phi}
depends only on the spatial part of $g_{ab}$ and is independent of
$N_A$. We also emphasize that the action (\ref{Cact1}) is finite,
and that it provides a valid variational principle for the above
boundary conditions for any $N_A(t_A)$.
Furthermore, given an action of the form (\ref{Cact1}), stationarity
of the action for fixed (conformal) boundary metric $g_{(0)CD}$ {\em
requires} this metric to be of the form (\ref{g0}), in particular
fixing the relationship between $g_{(0)CD}$ and the fixed $N_A(t_A)$ in (\ref{Cact1}). However, for now we take $N_A =1$ so that the boundary conditions
are manifestly $t_A$-translation invariant. 

Since the spin travels into the bulk at time $t_1$, it might appear
that Alice can no longer access the desired qubit after this
time. Such a conclusion would hold in a local non-gravitational
theory. But gravity changes this conclusion since both $\Phi_A(t_2)$
and ${\cal O}(t_2)$ are accessible to Alice at any time $t_2$.   As
a result, she needs only to measure $ e^{-i\Phi_A (t_2-t_1)} {\cal
O}(t_2) e^{i\Phi_A (t_2-t_1)} = {\cal O}(t_1)$. Here we have used
the fact (briefly reviewed in appendix \ref{manyH}) that $\Phi_A$ is the on-shell generator of
$t_A$-translations for $N_A =1$.

Now, to the extent that the bulk metric is in a semi-classical state with a well-defined causal structure\footnote{As noted in the introduction, since we are concerned with large, weakly curved regions of spacetime, one expects quantum fluctuations of the causal structure to be small.}, Alice can choose $t_2$ to be spacelike separated from the
event where Bob measures the qubit of interest.  This situation may
seem to give rise to a paradox.   On the one hand, since Alice is
just measuring ${\cal O}(t_1)$, it seems clear that the effect of
Alice's measurement must be identical to what would have occurred if
she had measured the qubit directly at time $t_1$. Such a
measurement would have correlated ${\cal O}(t_1)$ (say, the
$z$-component of a spin) with Alice's measuring device, so that  Bob
would receive the spin in what was effectively a mixed state.
Even if the spin was in a $S_x$-eigenstate before $t_1$, Bob would
find equal probability for both $S_x$-eigenstates when the spin
reaches his boundary.  On the other hand, Alice's measurement
occurred at time $t_A = t_2$, which by construction was spacelike
separated from Bob's experiment.  So, how did this decoherence
occur?

Answering this question requires a model of the couplings Alice engineers to perform her experiment; i.e., of the relevant modifications to (\ref{Cact1}).  Recall that Alice wishes to couple to $ e^{-i\Phi_A
(t_2-t_1)} {\cal O}(t_2) e^{i\Phi_A (t_2-t_1)}$.  Since the action is
a function of c-number field histories, it is not natural to include
such a commutator directly in the action.  However, the same effect is achieved by
modifying the action in three steps:

\begin{enumerate}[i)]

\item{}  At time $t_2 - \epsilon$ for small $\epsilon$,  add a term
$- \delta(t_2 - \epsilon -t_A) \ \Phi_A (t_2-t_1)$ to the
Hamiltonian; i.e., add $\int dt_A  \delta(t_2 - \epsilon -t_A) \
\Phi_A (t_2-t_1)$ to the action.

\item{} At time $t_2$, couple Alice's apparatus to the new ${\cal O}(t_2)$ so that she measures this observable.

\item{} At time $t_2 + \epsilon$, add a term
$- \delta(t_2 + \epsilon - t_A) \Phi_A (t_2-t_1)$ to the
Hamiltonian; i.e., add $\int dt_A \delta(t_2 + \epsilon - t_A) \
\Phi_A (t_2-t_1)$ to the action.
\end{enumerate}

The point of steps (i-iii) is that with these new couplings we have
\begin{equation}
\label{Osol}
{\cal O}(t_A) = e^{i \Phi_A f(t_A)} {\cal O}(t_1)  e^{-i \Phi_A f(t_A)} ,
\end{equation}
 where $f(t_A) = t_A - t_1 -(t_2 - t_1)  \chi_\epsilon(t_A) $ and $\chi_\epsilon$ is the characteristic function on the interval $|t_A - t_2| < \epsilon$; i.e.,  $\chi_\epsilon = 1$ for $|t_A - t_2| < \epsilon$ and  $\chi_\epsilon = 0$ for $|t_A - t_2| > \epsilon$.  In particular, step (ii) now measures ${\cal O}(t_2) = {\cal O}(t_1)$ as desired.  That (\ref{Osol}) is the correct solution is manifest from the relation
\begin{equation}
\label{otime}
\frac{d{\cal O}}{dt_A}(t_A) = i [\Phi_A f'(t_A), {\cal O}(t_A) ] = i  [H_A(t_A), {\cal O}(t_A) ],
\end{equation}
where $H_A(t')$ is the time-dependent Hamiltonian defined by steps (i-iii).\footnote{In the last equality of (\ref{otime}), we have used the fact that step (ii) adds a term to the Hamiltonian proportional to $\delta (t -t_2) {\cal O}(t_2)$.  Since this term commutes with ${\cal O}(t_2)$ and vanishes for $t \neq t_2$, it does not affect the evolution of ${\cal O}(t_A)$.}

We will need to analyze only step (i) in detail.  Because it subtracts a term from the Hamiltonian, we refer to this experiment as the $\Phi$-subtraction protocol.  Now, due to the
observations after eq. (\ref{phi}), adding the specified term to the
action is completely equivalent to shifting the lapse on boundary A
by $N_A \rightarrow 1 - \delta(t_2 -\epsilon - t_A) \  (t_2-t_1)$.
Thus, $N_A$ becomes a function of $t_A$ which in particular must
become {\em negative}.  This can also be seen in the fact that $f'(t_A)$ becomes negative in (\ref{otime}).  Even if the delta-function is replaced by a
smooth approximation, the lapse must still change sign and, in the
smooth case, must pass through zero.  Such boundary conditions are
incompatible with smooth invertible metrics, and any attempt to
define the theory requires input beyond our usual notion of
semi-classical gravity; i.e., we learn that the desired experiment cannot be described within the framework we have been using thus far.

It is of course an open question whether such boundary conditions
can be described in non-perturbative quantum gravity.  We will
discuss this issue in section \ref{disc} taking some input from
AdS/CFT.  However, having assumed that Alice has the ability to add
arbitrary couplings (and in particular the one associated with step
(i)), for now we simply suppose that such couplings are allowed and
press onward with our discussion.

We must therefore supply the required additional dynamical input by
hand. We shall do so using a certain analytic continuation.   To
begin, consider a less drastic version of steps (i-iii) associated
with an A-boundary lapse $N_A = 1 - \delta N_A(t)$, where this time
we take $\delta N_A(t) < 1$. In this case the analogues of steps
(i-iii) above merely implement a measurement of ${\cal O}$ at what for $N_A=1$ have been called time $t_2 - \Delta t$, where $\Delta t = \int \delta N_A(t)$.  The shift $N_A \rightarrow 1 - \delta N_A(t)$ is
essentially a change in the relationship between proper time on
boundary A and the time $t_A$ which governs the behavior of Alice's
ancilla, including any clocks present in Alice's laboratory.

It is therefore natural to suggest that the effect of (i-iii) above can be obtained by
analytic continuation in $\Delta t$: we declare that the net effect
of the original steps (i-iii) is equivalent to Alice simply
measuring ${\cal O}(t_1)$ directly at time $t_1$ except that, due to the above shift, the
relevant information appears in her measuring device only at time
$t_2$. In particular, although Alice's measurement occurs at $t_A =
t_2$ and thus would appear to have been causally separated from
Bob's measurement, the fact that Bob's measurement occurs in the
causal future of time $t_A =t_1$ nevertheless allows it to be
influenced by Alice's. Alice's experiment has fundamentally altered
causality in this system.

\section{A more physical measurement}

\label{projections}

The $\Phi$-subtraction protocol of section \ref{past}  involved couplings designed to allowed Alice
to recover information apparently sent into the bulk at a much
earlier time.  While these couplings may strike some readers as
rather contrived, the discussion served to illustrate a fundamental point:  Coupling directly to
the gravitational flux $\Phi_A$ changes the boundary conditions, and strong such couplings (of the correct sign) are incompatible with smooth invertible boundary metrics. Furthermore,  if the system can in fact be defined with such boundary conditions, one expects the effective
causal structure to be radically altered.

Since it is precisely the inclusion of $\Phi_A$ that makes the
algebra of A-boundary observables complete at each time, one might
expect this to be a generic feature of Alice's attempts to
holographically recover information at time $t_2$ which was
previously present on the A-boundary at time $t_1$.  Below, we
investigate this conjecture by analyzing a somewhat more physical
experiment in which Alice simply performs non-demolition
measurements of $\Phi_A$, ${\cal O}$, and $\Phi_A$ again in quick
succession. We refer to this experiment as the $\Phi$-projection protocol.  As will be explained in detail below, if her
measurements are of sufficient accuracy, and if she repeats such
measurements on a large number of identically prepared systems, she
can recover information associated with the operator ${\cal O}(t)$
any earlier time $t_2 - \lambda$. However, such experiments raise
issues quite similar to those of section \ref{past}.  The key point
is that any direct measurement of $\Phi_A$ involves a coupling to
$\Phi_A$, and that measuring $\Phi_A$ to high accuracy requires a
coupling that is in some sense strong.

To be specific,  consider a model in which Alice has 4 distinct
ancilla systems. The first is simply a spin, i.e., a $j= 1/2$
representation of SU(2). The associated SU(2) generators will be
denoted $S_x,S_y,S_z$ and we assume the spin to be prepared in the
$S_z = +1/2$ state.  The other ancilla are 3 pointer variables
described by canonical pairs $X_i,P_i$ (with canonical commutation
relations) for $i=1,2,3.$  These ancilla are initially prepared in
Gaussian wavepackets of widths $\sigma_i$ centered about $X_i =0$. For simplicity we take
all ancilla operators to be independent of time except as dictated
by their couplings to the AdS space; i.e., the free Hamiltonians of
Alice's ancilla vanish.  We again take the A-boundary metric to be
\eqref{g0} with $N_A =1$, except as modified by Alice's experiment
below.

 We model Alice's non-demolition measurements by
von-Neumann couplings \cite{vonN} to the pointer-variables
$X_1,X_2,X_3.$  The spin will be used to produce certain important
interference terms in the final stage of the experiment.  In
particular, although the spin is prepared in a spin up state (with definite $z$-component $S_z = +1/2$),
Alice will design her measurements to take place only if the $x$-component of the spin satisfies $S_x = +
1/2$.  At the end of the experiment, Alice measures the probability
that the spin and pointer-variables take various values.  The
resulting interference terms between the $S_x = \pm 1/2$ states will
allow her to determine $A(E,\lambda,E') := \langle \Psi|  P_{\Phi_A
= E} P_{{\cal O}(t_2) = \lambda} P_{\Phi_A = E'} |\Psi \rangle$
where $|\Psi \rangle$ is the quantum state of the system (see
footnote \ref{foot}).  The probability distribution of ${\cal
O}(t_1)$ may then be recovered by integrating $A(E,\lambda,E')$
against $e^{-i(E-E')(t_1-t_2)}$.  As usual in quantum mechanics,
Alice must have access to arbitrarily many identically prepared
copies of the AdS space to measure the above probabilities.  We
assume that this is the case.

The details of the $\Phi$-projection protocol can be described in the Schr\"odinger
picture as a sequence of unitary transformations and projections
onto apparatus variables.  The procedure is:

\begin{enumerate}[{\bf i:}]

\item {} Apply $\exp\left(ig_1 \Phi_A (S_x + 1/2) P_1\right)$.  If $S_x = +1/2$, this implements a von Neumann measurement of $\Phi_A$ by $X_1$ with coupling
$g_1$.

\item {} Apply $\exp\left(ig_2 {\cal O} (S_x + 1/2) P_2\right)$.  If $S_x = +1/2$, this implements a von Neumann measurement of ${\cal O}$ by $X_2$ with coupling
$g_2$.

\item {}  Apply $\exp\left(ig_3 \Phi_A (S_x + 1/2) P_3\right)$.  If $S_x = +1/2$, this implements a von Neumann measurement of $\Phi_A$ by $X_3$ with coupling
$g_3$.

\item {} Choose parameters $x_1,x_2,x_3$ and apply $\exp\left(-i(S_x - 1/2) (x_1P_1+ x_2P_2 + x_3P_3)\right)$.  If $S_x = -1/2$ (so that none of the above measurements took place), this translates $X_1,X_2,X_3$ by $x_1,x_2,x_3$.

\item {}  Project onto eigenstates of $X_1,X_2,X_3$ with
eigenvalues $x_1,x_2,x_3$ (more properly, onto corresponding
spectral intervals); i.e., measure the operators $X_1,X_2,X_3$ and
abort the experiment unless the same values are obtained as were chosen in step (iv).

\item {}  Choose a unit vector $\vec v \in {\mathbb R}^3$
and project onto states with $\vec v \cdot \vec S = +1/2$; i.e.,
measure  $\vec v \cdot \vec S$ and abort the experiment unless
the values $+1/2$ is obtained.
\end{enumerate}

By the usual rules of quantum mechanics, the probability that the experiment succeeds (i.e., that the
experiment is not aborted in either stage (v) or stage (vi)) is given by
\be
 \label{P123}
 P(x_1,x_2,x_3,\vec v) = \frac{1}{2}  \Bigg|  \alpha |\Psi \rangle +
 \beta P_{H_A=x_3} P_{{\cal O} = x_2}
P_{H_A=x_1} |\Psi \rangle  \Bigg| ,
 \ee
where, with appropriate conventions for the spin-eigenstates, we have
 \be
 \alpha = i \langle \vec v \cdot \vec S = +1/2 | S_x = -1/2
\rangle, \ \ \  \beta = \langle \vec v \cdot \vec S = +1/2 | S_x = +
1/2 \rangle.
 \ee
By repeating the experiment many times on identically prepared systems and varying the choice of $x_1,x_2,x_3, \vec v$, Alice can determine the entire function (\ref{P123}) to arbitrary accuracy.  Note that $|\alpha|^2 + |\beta|^2 =1$, but that $\alpha$ and
$\beta$ may otherwise be chosen arbitrarily. From her
measurements of $ P(x_1,x_2,x_3,\vec v)$, Alice may thus calculate the
term in \eqref{P123} proportional to $\alpha^* \beta$; i.e., she may
calculate the amplitude
 \be
 \label{A123}
 A(x_1,x_2,x_3,\vec v) = \langle \Psi |  P_{H_A=x_3} P_{{\cal O} = x_2}
P_{H_A=x_1}) |\Psi \rangle.
 \ee
The probability distribution of ${\cal O}(t_2 - \lambda)$ may be then
recovered by integrating \eqref{A123} against $e^{-i\lambda x_1}
e^{i \lambda x_3}.$  Similarly, any other data that Alice might have
accessed at time $t- \lambda$ can be accessed at time $t$ by simply
replacing step (ii) with the procedure  to measure this data
directly, conditioned as above on having $S_x = + 1/2.$

As in section \ref{past}, we wish to understand the impact of
Alice's measurements on dynamics, and in particular on the boundary conditions.   Each step in the $\Phi$-projection protocol is of course associated with the addition of an
appropriate term to the action.  The terms of most interest will be
those associated with steps (i) and (iii) which take the form
 \be
\label{i+iii} S_{(i)+(iii)} =  - \int dt_A \Bigl( f_1(t_A) \Phi_A
(S_x + 1/2) P_1 + f_3(t_A) \Phi_A (S_x + 1/2) P_3 \Bigr),
 \ee
where $\int dt_A f_1(t_A) = g_1$ and  $\int dt_A f_3(t_A) = g_3$.
Such terms resemble the couplings of section \ref{past} with the
magnitude of the coupling being set by $f_1(t_A)(S_x + 1/2)P_1$ and
$f_3(t_A)(S_x + 1/2)P_3$.

When $f_3(t)=0$,
the boundary term \eqref{i+iii} forces the
A-boundary lapse to be $N_A = 1 - f_1(t_A) (S_x + 1/2) P_1$.  Since the case of interest is $S_x = +1/2$, the lapse remains positive  only if $f_1(t_A) P_1 < 1$.  Imposing such a
requirement would restrict the resolution of the
measurement in terms of the time $\Delta t_A$ which elapses during
the experiment.  In particular, it would require $g_1 \Delta P_1 < \Delta
t_A$, where $\Delta P_1 = 1/\sigma_1$ is the momentum-space width of
the Gaussian initial state for this pointer-variable.  Since the
position-space width is $\Delta X_1 = \sigma_1$, and since the
interaction translates $X_1$ by $g_1 \Phi_A$, Alice's experiment
measures $\Phi_A$ with a resolution $\Delta \Phi_A =  \frac{1}{g_1
\Delta P_1}$.  Keeping the lapse positive would thus require $\Delta
\Phi_A > \frac{1}{\Delta t_A}.$  While this is reminiscent of an
energy-time Heisenberg uncertainty relation, it is important to
recall that other quantum systems do allow better measurements of
energy on much shorter timescales \cite{AB}. We will save for
section \ref{disc} any discussion of whether $\Delta \Phi_A \Delta
t_A > 1$ constitutes a fundamental restriction in the AdS context or
merely limits the familiar semi-classical framework.

Now, how accurately does Alice need to measure $\Phi_A$ in order to
recover information at $t_A = t_1$?  If she makes no assumptions
about the spectrum of $\Phi_A$, she must allow for frequencies of order
$\frac{1}{t_2-t_1}$, where $t_2$ is the time at
which stage (ii) is performed.  Alice thus needs $\Delta \Phi_A \sim
\frac{1}{t_2-t_1}$ to obtain even rough information, and she will
require $\Delta \Phi_A \ll \frac{1}{t_2-t_1}$ to obtain high
resolution.  But if $t_1$ occurs before the experiment begins, then
since stage (i) itself takes time $\Delta t_A$ we have $t_2 - t_1 >
\Delta t_A$.  Thus $\Delta \Phi_A  \Delta t_A \ll 1$ for a precision
measurement. In summary, if she makes no assumptions about the
spectrum of $\Phi_A$, obtaining significant information about
observables before her experiment began requires Alice to use
couplings strong enough to raise the same issues as in section
\ref{past}.  Again, if we assume that such couplings are
nevertheless allowed, the natural conclusion is that they alter the
naive notion of causality so that Alice's experiment can effect
Bob's.  While Alice measures a coherent qubit, the qubit Bob
receives is in a mixed state as if the $z$-component of its spin had
already been measured.

\section{Operationally discrete spectra}
\label{discrete}

Section \ref{projections} discussed the $\Phi$-projection protocol making no assumptions about the spectrum of $\Phi_A$.  Of course, it is also interesting to suppose that Alice {\em does}
know something about the spectrum of $\Phi_A$.  An interesting case
arises if this spectrum is discrete, so that any resolution finer
than the smallest level spacing suffices to obtain information about
the very distant past.  Thus, Alice may be able to complete her
measurement using couplings compatible with familiar AdS asymptotics and
avoiding radical effects on the causal structure.

In fact, we will require finiteness only of the A-boundary's
`operational density of states.'   The idea is that only states which
can be actively probed from boundary A are relevant, and that we
discard any other states in computing this density. After
introducing this notion below, we reconsider the $\Phi$-projection protocol in section \ref{return}.  We also consider a new experiment (the quantum computer protocol) in section \ref{generic} which does not involve direct couplings to $\Phi_A$.

To define Alice's operational density of states, we first suppose
that Alice has access to a large number of AdS systems which define
identical states $\rho$ on the A-boundary observables. We explicitly
allow $\rho$ to be a mixed state and use the notation of density
matrices.  We emphasize that only the restriction of the state to
A-boundary observables is relevant, and that these states need not
be identical in any deeper sense.

Now consider the Hilbert space defined by the Gelfand-Naimark-Segal
construction (see e.g. \cite{Dixmier}) using $\rho$ and this
observable algebra; i.e., for each (bounded) observable ${\cal O}_A$
at Alice's boundary we define a state $| {\cal O}_A \rangle$ and
introduce the inner product

 \be
 \langle {\cal O}_A^1 | {\cal O}_A^2 \rangle = Tr \left( \rho ( {\cal O}_A^1)^\dagger {\cal
 O}_A^2 \right).
 \ee
The right-hand side is positive semi-definite and sesquilinear.  We
may thus quotient by the zero-norm states and complete to define
Alice's `operational' Hilbert space ${\cal H}_A$. We take her
operational density of states to be the entropy $S(E)$ defined by
the operator $\Phi_A$ on ${\cal H}_A$.   If $S(E)$ is finite, we say
that the density of AdS states is operationally finite. In cases
where some AdS states cannot be distinguished by A-boundary
observables, the true number of states can be far larger than
$S(E)$.

The entropy $S(E)$ counts the density of states with $\Phi_A = E$
that can be distinguished using A-boundary observables.  It is thus
tempting to use the gravitational thermodynamics of asymptotically AdS
spaces to conclude, at least in the absence of an inner boundary,
that $S(E)$ {\em must} be finite and that for large $E$ it is given
by the AdS Bekenstein-Hawking entropy $S_{BH}(E)$.  This conclusion
will hold if time-independent couplings of the AdS system to Alice's
finite-entropy ancilla generically lead to thermodynamic equilibrium
states in which the AdS system is well-described by semi-classical
calculations.    However, we saw in sections \ref{past} and
\ref{projections} that strong couplings to $\Phi_A$ take us outside
the usual framework of semi-classical gravity.  Thus, this framework
cannot be said to probe generic couplings.  We will return to this
issue in section \ref{disc}, but for the rest of this section we
simply assume that $S(E)$ is finite without imposing any
 particular restriction on its form.

The discussion above has not explicitly mentioned either Bob or any
 inner boundary.  If they are present,  Bob and his ancilla
are merely part of the system that Alice probes through her
couplings to the AdS boundary, and Alice need not distinguish them
from the bulk AdS system.    This is another reason not to specify
the form of  $S(E)$; this density {\em will} generically depend on
the ancilla that Bob couples to the AdS space.

Even just taking $S(E)$ to be finite imposes certain restrictions on Bob's
couplings.  In particular, it forbids most explicitly time-dependent
couplings at boundary B.   The point is that acting with $\exp(i \lambda \Phi_A)$
translates boundary A relative to boundary B.  As a result, if Alice
can send signals which probe Bob's measuring devices and return, and
if Bob's couplings determine a preferred time $t_0$ in the original
state $\rho$, the observables at boundary A are sensitive to $t_0 -
\lambda$.  Acting with $\exp(i \lambda \Phi_A)$ then generates an
infinite-dimensional Hilbert space of states distinguished by
A-boundary observables. One exception occurs when Bob's couplings
are periodic, though in that case any analysis is much like the
time-independent case.  One might also attempt to forbid Alice from
actively probing Bob's couplings, though it is not clear how this
can be done.   In particular, if the state $\rho$ was such that
Bob's couplings turned on only inside a black hole, then acting with
$\exp(i \lambda \Phi_A)$ can translate the system to a state where
the above $t_0$ occurs before the black hole formed or, for
classically eternal black holes, to when it experienced a rare
quantum fluctuation into a horizon-free spacetime filled with
thermal radiation.  One concludes that Bob's couplings are not truly
hidden and that the operational density of states will again diverge
if his couplings define a distinguished time $t_0$.

We therefore require Bob's couplings
to be time-independent below.  This makes sense only when the boundary
conditions at boundary B have a time-translation symmetry; i.e.,  for
Dirichlet-like boundary conditions the (fixed) metric on boundary B
must be stationary.  It is not immediately clear to what extent such
boundary conditions are compatible with the interesting case where
Bob enters (the future-trapped region of) a black hole. A proper
treatment of such cases may require more flexible boundary
conditions, and in any case is complicated by failure of classical
physics near the black hole singularity.  We therefore avoid this
setting in sections \ref{return} and \ref{generic} below, though we
provide a few brief comments in section \ref{BH}.

\subsection{A return to projections}
\label{return}

We now reconsider the $\Phi$-projection protocol of section
\ref{projections} under the assumption that the AdS system has an operationally
finite density of states for $\Phi_A$, and further assuming that
Alice knows the spectrum of $\Phi_A$ precisely. This may be either
because she has solved the full quantum theory, or because she has
already performed a large number of experiments to determine this
spectrum.

The typical spacing between $\Phi_A$-eigenstates is $\Delta \Phi_A
\sim \mu e^{-S(E)}$, where $\mu$ is an appropriate energy scale.
Thus, by allowing both stages (i) and (iii) to take time $\Delta t_A
\gg \mu^{-1} \exp(S(E))$, Alice can obtain accurate information
about $A(E,\lambda,E',\vec v)$ for essentially all eigenvalues
$E,E'$ of $\Phi_A$ while still satisfying $\Delta t_A \Delta \Phi_A
> 1$.  She can then use this information to extrapolate back to {\em
much} earlier times.  The the only errors in her calculation arise
from the off chance that she measured an eigenvalue $E_i$ for
$\Phi_A$ when the actual result was some other eigenvalue $E_j$.
Since we began with detectors in Gaussian wave packets $\propto
e^{-x_1^2 /\Delta x_1^2}$, the probability for this to occur is
Gaussian in $E_i - E_j$ and is typically of order $ \exp
\left(-g_1^2 \mu^2 e^{-2S(E)} /\Delta x_1^2\right) \sim \exp \left(-
\Delta t^2_A \ \mu^2 e^{-2S(E)} \right) $, where we have chosen
$f_1(t)$ such that $\Delta t_A \Delta \Phi_A \sim 1$. Since there
are $\exp(S(E))$ states, and since the full state enters
quadratically in Alice's calculation, her total error is of order
 \be
 \exp \left(2 S(E) - \Delta t^2_A \ \mu^2 e^{-2S(E)}
 \right),
 \ee
and so is exponentially small for $\Delta t_A \gg \frac{\sqrt{S(E)}
}{\mu} e^{S(E)}$.    Thus, provided that no energy levels have an
unnaturally small splitting of eigenvalues, for such $\Delta t_A$  there is
essentially no limit to Alice's lookback time.  It is interesting to note that the same conclusion also holds in the presence of exact degeneracies (e.g., due to symmetries); for our present purposes, there is no need to distinguish states with identical time-dependence.

Due to the long timescale $\Delta t_A$, it is not difficult to
reconcile Alice's measurements with Bob's measurement of a
non-commuting observable. We suppose that Bob arranges a
time-independent coupling to his devices at boundary B, and that
this coupling is consistent with the finiteness of Alice's
operational density of states $S(E)$. Such an interaction might be
triggered by the approach of spins with certain characteristics, but
the coupling remains non-zero at all times. Bob's device is a like a
photodetector that is always on.  As a result, while information may
flow into Bob's device during the experiment, the information can
leak back out if Alice allows her experiment to run for a long
enough time. Since $\Delta t_A \sim \exp \left( S(E) \right)$, any
information remaining in Bob's ancilla is associated with states
split in energy by much less than the typical value $e^{- S(E)}$ assumed above.  If such states
exist, they limit the success of Alice's experiment in precisely the
same way as would any other finely-tuned near degeneracies in the
spectrum of $\Phi_A$.  On the other hand, to recover the desired
information, there will be some timescale over which all information
does leak out of Bob's ancilla.  Alice simply needs to extend the
experiment to run over this longer period of time.

\subsection{Quantum computers and generic couplings}
\label{generic}

We noted above that an operationally finite density of states allows
Alice to perform useful holographic experiments without radical
alterations of the causal structure at her boundary.  The particular
experiment discussed used measurements over very long times $\Delta
t_A \gg e^{S(E)}$ to measure $\Phi_A$ to great accuracy. It is
therefore interesting to ask if similarly useful experiments can be
performed over shorter timescales or with more generic couplings.

We now argue that this is the case, and that  (at least when Bob
does not interfere) one should be able to reduce $\Delta t_A$ to
roughly the timescale associated with the evaporation of black holes
in flat space.     In this experiment, Alice will couple a small
quantum memory device ($QM_1$, with entropy $S_1 \ll S(E)$) to the
A-boundary in a fairly generic way, let the system equilibrate, and
then couple the A-boundary to a large quantum memory device ($QM_2$,
with entropy $S_2 \gg S(E)$) .  If $S_2$ is sufficiently large,
almost all of the information originally available in $QM_1$ will be
accessible from $QM_2$ once the system reaches its final
equilibrium.  The argument itself is not particularly novel:  we
merely use the idea that there is a unitary generator $H_A$ of
time-translations along the A-boundary to translate standard
reasoning to our setting from non-relativistic quantum mechanics. In
particular, we will make use of the fact emphasized in appendix \ref{manyH}
that the use of time-dependent couplings merely makes $H_A$ a
time-dependent function of A-boundary observables and Alice's
ancilla.

As before, we assume Alice's operational density of states to be
finite. However, for this new experiment, we also assume the system
Alice probes to have an `operationally unique ground state' (though
our argument readily extends to the case of multiple vacuua so long
as Alice can distinguish such vacuua by non-demolition experiments).
Our specific assumption is that, if Alice were to couple ancilla
with an infinite density of states  to the A-boundary, the system
generically relaxes to a state such that

\begin{itemize}

\item {} Alice's boundary observables are uncorrelated with any of
her other degrees of freedom.

\item {} The expectation value $Tr ( \rho {\cal O}_A )$ of any A-boundary observable is independent of both the coupling used and the initial state $\rho_i$ (so long as it is a density matrix on ${\cal H}_A$).

\end{itemize}
These assumptions again involve only the restriction of the state to
Alice's observables; we make no assumptions about any further
observables which might be inaccessible to Alice.

In general, one expects the above relaxation to be rapid compared
with the exponentially long timescales $e^{S(E)}$ of section
\ref{return}. Certainly, free radiation in AdS will propagate to
where it registers in A-boundary observables on timescales
comparable to the AdS scale.  Thus, such radiation can be rapidly
extracted by Alice's boundary couplings.  While the relaxation
proceeds more slowly in the presence of black holes, the couplings
can allow Hawking radiation to rapidly leak out through the AdS
boundary.  One therefore expects the relevant timescale to be some power law
in the energy resembling the timescale for black hole decay in flat
space\footnote{In fact, as noted in \cite{comp,key}, with certain additional assumptions (concerning either the form of $S(E)$ or the ``mixing time'), versions of this experiment with Bob inside a black hole may in fact be conducted over much shorter timescales, in some cases only logarithmically longer than the light-crossing time of the black hole.  However, for simplicity we avoid such extra assumptions below.}.  As a result, at least when Bob's ancilla are not coupled to
the system, one expects this experiment to proceed much faster than that of
section \ref{return}.

Assuming that the ground state of $QM_1$ is unique, the argument is
now immediate.  Alice couples first $QM_1$ and then $QM_2$ to
boundary A and lets the system equilibrate.  Both $QM_1$ and the
A-boundary observables are then in their ground states, and the
final state of $QM_2$ is unitarily related to the initial state of
$QM_1$. To see this,  one need only solve the Heisenberg equations
of motion at boundary A \eqref{Aevolve2} to relate any late time
operator ${\cal O}_{QM_2}$ of $QM_2$ to the early time operators of
$QM_1$, $QM_2$, and the observables at boundary A.  The algebra of operators defined by  $QM_1$, $QM_2$, and boundary A at an early time $t_A =t_1$ thus suffices to compute $Tr
(\rho {\cal O}_{QM_2})$ at any time.

Similarly, any observable of $QM_1$ at $t_1$ can be expressed
in terms of observables for $QM_2$, $QM_1$, and boundary A at any late
time $t_A = t_2$.  Since the A-boundary relaxes to a known state and $QM_1$ relaxes to its (known) ground state, correlators of early time operators for $QM_1$ can be computed in terms
of late-time correlators of $QM_2$; i.e., the full information in the
initial state of $QM_1$ can be recovered from the observables of $QM_2$.

So long as his couplings do not destroy the above assumptions,
including Bob requires no changes in this discussion.  As in section
(\ref{return}), his measurements are easily reconciled with those of
Alice.  Because he leaves all of his couplings turned on,
over the long time it takes Alice's experiment to run any
information in his ancilla can leak back out to the AdS boundary.
 It is true that if Bob's couplings are weak or if the entropy of his ancilla is
large, his presence can greatly affect the time required for the
A-boundary to relax to its ground state (and thus for equilibrium to
be reached).  However, since Alice has access to arbitrarily many
identical copies of the AdS system (coupled identically to Bob's
ancilla), she may simply measure this relaxation time and then
design her experiment accordingly.

\subsection{Experiments inside Black Holes}
\label{BH}

Perhaps the most interesting setting for our experiments occurs when
Bob (or, more properly, boundary B) falls into a black hole.
However, as noted earlier, it is unclear to what extent such situations
are consistent with time-translation invariance at
boundary B, and in particular with taking the metric on boundary B
to be stationary, which was assumed for all experiments in this section (the weak coupling $\Phi$-projection protocol and the quantum computer protocol).

Nonetheless, since the experiments above last long enough for any
black hole to either evaporate or to fluctuate into a horizon-free
geometry, the details of Bob's experience inside the black hole may
not be relevant.  Suppose, for example, that boundary B remains
present after the black hole evaporation or fluctuation, and that it
remains connected to the same asymptotic region of spacetime.  In
that case the discussions above continue to apply, though with new
details that may be of interest.

Let us examine these details in the context of the
quantum computer protocol (section \ref{generic}).  Recall that
Alice couples only to outgoing radiation, which may consist both of
Hawking radiation and of additional radiation emitted by Bob's
ancilla after the evaporation of the black hole.    In the absence
of boundary B, unitarity would imply that the von Neumann entropy of
the Hawking radiation is the same as that of the state which formed
the black hole.  The mechanism for this was outlined in
\cite{paper1}, and the key step was to relate the A-boundary
Hamiltonian to the Hawking radiation. As the black hole evaporates,
one notes that the gravitational Gauss' law relates the radiation
stress tensor to the difference between $\Phi_A$ and the
corresponding gravitational flux $\Phi_{horizon}$ at the black hole
horizon. When the horizon disappears, $\Phi_{horizon}$ vanishes and
$\Phi_A$ is completely encoded in the Hawking radiation.

However, if boundary B remains present after evaporation, the
gravitational Gauss' law relates $\Phi_A$ to both the radiation stress
tensor and to a similar gravitational flux $\Phi_B$ at boundary B.
The von Neumann entropy of the Hawking radiation thus remains linked
to that of Bob's ancilla through $\Phi_B$.   Until Bob's ancilla
spontaneously de-excite and decorrelate themselves with the bulk AdS
space, the A-boundary observables will not relax to their ground
state.  Alice's experiment must run for a time dictated by Bob's
ancilla and not just by Hawking evaporation of the black hole.
Similar conclusions can be reached for the $\Phi$-projection protocol of section
\ref{return}.

In contrast, one might also investigate the case where boundary B ceases
to exist after evaporation of the black hole.  Versions of the quantum computer protocol were studied for such cases in \cite{comp,key,scramble}.    Due to making additional, assumptions about either $S(E)$ or the ``mixing time,''  refs. \cite{comp,key,scramble} considered experiments that ran for much shorter times than ours, though such times were always at least logarithmically longer than the light-crossing time of the black hole.   We have nothing new to add to this discussion here and continue to rely on the resolution suggested in \cite{comp,key,scramble}.  In particular, since the quantum computer protocol couples directly to the Hawking radiation, it is difficult to see how it could lead to causality-violating effects of the sort caused by our short-time $\Phi$-subtraction and $\Phi$-projection protocols.  Instead, \cite{comp,key,scramble} argued that  no true paradox could result unless the observers were able to compare the results of their experiments, and that the time required for these experiments was long enough to make comparison impossible before Bob is destroyed in the black hole singularity.

Finally, one might consider cases where boundary B continues to exist beyond the black hole singularity, but where  it ceases
to be connected to the same asymptotic region.  Perhaps it enters a
`baby universe.'  In such cases it is more difficult to reconcile
Alice and Bob's non-commuting measurements, though this might be
possible in some more complete theory.  If not, then baby universe
production may be incompatible with an operationally finite density of
states (and with an operationally unique ground state).

\section{Discussion}
\label{disc}

We have explored a number of thought experiments in asymptotically
AdS quantum gravity featuring holographic measurements performed
by a boundary observer (Alice). Our focus was on experiments in which Alice couples directly to the gravitational flux $\Phi$ associated with the boundary term in the gravitational Hamiltonian, as opposed to attempts to extract information directly from outgoing radiation.  We also allowed for a second
observer (Bob) who performs a more local measurement.  Both
observers were taken to lie outside the spacetime so that there was
no danger of Alice having access to a holographic encoding of Bob,
and so that we could cleanly discuss gauge-invariant observables.
The goal was to make more concrete the notion of boundary unitarity
discussed in \cite{paper1} and to resolve various potential
paradoxes.  It is clearly also of interest to understand the extent
to which holographic measurements are possible for observers who are
themselves part of the gravitating system, but we have not pursued
this question here.

Interesting cases arise when the two observers measure operators
that do not commute. The first class of settings (sections
\ref{past} and \ref{projections}) seemed particularly paradoxical as
the measurements occured at events which, in the absence of the
measurements, would not have been causally connected. But by general
principles of quantum mechanics, non-commuting measurements should
interfere with each other. Moreover, Alice's holographic
measurements were guaranteed to succeed as planned under the
assumptions of \cite{paper1}.  Thus, it was Alice's holographic
measurement which must somehow interfere with Bob's familiar local
measurement, despite the apparent causal structure.

The resolution was that, for each experiment,  a complete analysis
was not possible within the usual framework of semi-classical
gravity.  Furthermore, the particular form of this failure suggested
radical modifications to the naive causal structure.  In particular,
these experiments involved strong couplings to the gravitational
flux $\Phi_A$ associated with the usual Arbowitt-Deser-Misner (ADM)-like boundary term in
the Hamiltonian.  Such couplings were shown to alter the boundary
conditions in a manner incompatible with smooth invertible metrics,
even at the asymptotic boundary.  Instead, they required the lapse
$N_A$ at this boundary to pass through zero and become negative.  We
argued by analytic continuation that, {\it if this behavior is allowed} in
the full theory of AdS quantum gravity, we expect it to modify the
causal structure so that Alice's experiment can in fact influence
Bob's. In the scenarios discussed, Alice's measurement proceeded as
she expected but resulted in Bob receiving what was effectively a
mixed state.  I.e., the result was the same as if Alice's
measurement had occurred in Bob's causal past.

Given that they force us out of the familiar semi-classical domain,
the reader may wonder whether the couplings of sections \ref{past}
and \ref{projections} (the $\Phi$-subtraction and $\Phi$-projection protocols) are in fact allowed in any complete theory.
Could it be that we have granted Alice unphysical powers in making
her measurements, perhaps in the same way that certain measurements
are unphysical in relativistic field theory \cite{Sorkin,Preskill}?
Since a complete answer requires some input from quantum gravity, it
is enlightening to ask this question in the context of AdS/CFT:
Suppose that the AdS system has a dual formulation in terms of some
large N gauge theory, and that it is this gauge theory which sits in
a box in Alice's lab.  In that context, we see no obstacle to making
precise measurements of the energy on short timescales.
In particular, recall that Aharonov and Bohm showed \cite{AB} how,
for non-relativistic quantum systems, precise measurements of energy can be made arbitrarily
rapidly.  In the relativistic case, one expects that any additional restrictions are
set by the light-crossing time of the gauge theory system in Alice's
laboratory and not by the intrinsic resolution of the measurement.
Thus, at least in this context, the $\Phi$-projection experiment of section
\ref{projections} seems to be allowed.

The second class of settings (section \ref{discrete}) was less
intrinsically paradoxical, but maintained the standard causal
structure on the boundary.   In such settings,  Alice's experiments
lasted for long enough intervals of time to place Bob and Alice in a
form of causal contact\footnote{Though in some cases this required
the evaporation of black holes or their fluctuation into
horizon-free geometries, in which case we had to make further
assumptions about how this affected Bob's boundary.  See section
\ref{BH}.}. However, these experiments succeed only if the AdS space
has an `operationally finite density of states' $S(E)$.  We noted
that the details of both $S(E)$ and the timescale the experiment
requires may depend on Bob's choices of ancilla and couplings.

The discussion above allowed Bob to work at a finite boundary, at
finite distance from bulk events.   Suppose however that we imposed
more familiar boundary conditions allowing only asymptotic
boundaries. Since we know of no classical solutions in which two
asymptotically AdS boundaries are causally connected, it is natural
to assume that the A-boundary density of states $S(E)$ is
independent of any ancilla or couplings at other boundaries.  In
this context, one might hope to calculate $S(E)$ from semi-classical
gravitational physics, and it is tempting to conclude that it agrees
with the Bekenstein-Hawking entropy $S_{BH}(E)$ at large $E$.  In
particular, we note that $S(E)$ is precisely the density of states
that can affect the exterior of the black hole, which was advocated
to correspond to black hole entropy in e.g. \cite{Banks,Ted}.  One
possible loophole is that some dynamical selection mechanism might
forbid certain states described by $S(E)$ from appearing in thermal
equilibrium, and it was noted in section \ref{discrete} that this
might occur if high resolution measurements of $\Phi$ are
fundamentally forbidden.  However, we have now argued that such
measurements are allowed (at least in the context of AdS/CFT),
making this loophole less plausible.

While our discussion above was cast in terms of effects on the causal structure due to the influence of Alice's experiments, the reader may wonder if quantum fluctuations of the causal structure play any role.  On the one hand, as noted in the introduction, we are largely concerned with weakly curved regions of spacetime near the AdS boundary where one would expect such quantum fluctuations to be small. On the other hand, since the causal structure is a dynamical variable, it does not generally commute with the Hamiltonian (i.e., with $\Phi$).  As a result, at least in the interior of the spacetime, one might expect measurements of $\Phi$ with small uncertainty $\Delta \Phi$ to lead to large fluctuations in the causal structure, and one might further attempt to interpret our results in these terms.  However, recall that section \ref{projections} found no tension between precise measurements of $\Phi$ and a well-defined asymptotic causal structure, so long as the measurement was carried out over a sufficiently long time.  This argues against the existence of any simple energy-causal structure uncertainty relation that could replace our analysis above.  It would, however, be interesting to analyze the relevance of quantum causal structure fluctuations in more detail.

As a final remark, the reader should note that the resolutions
described above are quite different from those proposed in
\cite{comp,key,scramble} for related thought experiments.  Because they
studied the extraction of information from Hawking radiation, and
because the observer outside the black hole had to wait long enough
to collect enough radiation, these works found that the two
observers were unable to compare their results after the experiments
were completed. The authors argued that, as a result, no true
paradox could arise. In contrast, our settings include those where
the observers can compare results. In particular, we considered short-time versions of the $\Phi$-subtraction and $\Phi$-projection protocols in sections \ref{past} and \ref{projections}.  Whether or not comparison is possible, our main conclusion was that a sufficiently accurate holographic measurement necessarily causes the boundary metric to degenerate, taking us out of the realm of familiar gravitational physics.  In contexts such as AdS/CFT where these high-resolution experiments are nevertheless allowed, we argued that it leads to a radical change in the effective bulk causal structure.  The result is that the holographic experiment {\it can} affect results obtained by an a priori causally separated second observer deep in the interior, so that this second (internal) observer receives a state already decohered by the holographic measurement.  Thus the internal observer effectively receives a mixed state from which no paradoxes can arise.

\subsection*{Acknowledgements}

The author has benefited from many discussions with physicists at
UCSB, the Perimeter Institute, the ICMS workshop on Gravitational
Thermodynamics and the Quantum Nature of Space Time in Edinburgh,
and the ICTS Monsoon Workshop on String theory in Mumbai. In
particular, he thanks Roberto Emparan and
especially Ted Jacobson and Simon Ross for their
thought-provoking comments and questions. This work was supported in
part by the US National Science Foundation under Grant
No.~PHY05-55669, and by funds from the University of California. The
author thanks the Tata Institute for Fundamental Research and the
International Center for Theoretical Sciences for their hospitality
and support during critical stages of this project.

\appendix

\section{Diffeomorphism invariance and the Hamiltonian}

\label{manyH}

This appendix provides a brief reminder of certain technical
details associated with charges and symmetries in
diffeomorphism-invariant theories.  We wish to address three sorts
of complications: i) situations with multiple boundaries, ii) the
coupling of external (non-gravitating) degrees of freedom to
boundary observables and iii) time-dependent boundary couplings
(i.e., time-dependent boundary conditions).  Situations of interest
will typically involve all three issues simultaneously.  Our treatment of time-dependent boundary conditions below will be fairly formal.  In contrast, appendix \ref{exB} examines a particularly simple example of time-dependent couplings between a bulk (scalar) field and an external system in detail.   As a result, readers seeking physical insight into such time-dependent couplings are advised to first read  appendix \ref{exB}.

The general setting for our discussion is an action functional
defined on a gravitating system (with boundaries) as well
as some additional degrees of freedom (ancilla) associated
with each boundary.   For definiteness and simplicity, let
us consider the case of two boundaries (A,B) which is of most
interest in the main text.  These may be either finite boundaries (in which the boundary lies at finite proper distance from the interior) or conformal boundaries with AdS asymptotics.

The ancilla associated with boundary A (B) are denoted $\alpha_A$
($\alpha_B$). On each boundary (A,B) we choose some time coordinate
$(t_A,t_B)$  (such that the surfaces $t_A= constant$, $t_A=
constant$ are Cauchy surfaces within the respective boundaries)
which will define a notion of causality respected by the ancilla.
The action will be stationary under an appropriate boundary
condition which relates the ancilla $\alpha_A, \alpha_B$ to the
fields and their derivatives on a finite boundary, and to the
Fefferman-Graham coefficients (see e.g.  \cite{FG,KSthermo}) of the
bulk fields at an AdS conformal boundary.  Below, we use the term
`boundary values' to refer to both the fields and their normal
derivatives at a finite boundary, and to the two independent
Fefferman-Graham coefficients for each field at an AdS conformal
boundary.   What is important for our purposes is that these
boundary conditions may be chosen to share any symmetries of the
action, and that the boundary conditions break diffeomorphism
invariance (so that boundary diffeomorphisms are not gauge
symmetries).  In particular, we assume that all boundary values of
bulk fields are gauge-invariant observables.

We assume the action to be invariant under diffeomorphisms generated by
vector fields that vanish sufficiently rapidly at the (perhaps
conformal) boundaries of the spacetime (see e.g. \cite{KSthermo} for AdS details).
We take the entire action to be the integral of a local density over
the bulk spacetime, an appropriate set of (local)
boundary terms which depend only on boundary values of bulk fields, and two additional terms of the form
 \begin{equation}
 S_{int} = \int dt_A  L_A  + \int dt_B L_B,
 \end{equation}
where $L_A$ ($L_B$) is a function of both the $\alpha_A$ ($\alpha_B$) and
the A-boundary (B-boundary) observables at time $t_A$ ($t_B$). Any
coupling functions appearing in $L_A$ ($L_B$) are allowed to depend
only on the time coordinate $t_A$ ($t_B$). Thus $S_{int}$ describes
the full physics of the ancilla, including any interaction terms.

Let us first suppose that the action does not explicitly depend on
$t_A$, and that the boundary vector field $\frac{\partial}{\partial t_A}$ can be smoothly extended into the bulk such a way that the diffeomorphism it generates preserves both the action and boundary conditions.  Because diffeomorphisms that vanish sufficiently rapidly at the
boundaries are pure gauge, this means that the action is invariant
under the simultaneous transformations $t_A \rightarrow t_A +
\tau$ on the ancilla $\alpha_A$ and a diffeomorphiism of the
AdS space which restricts to $t_A \rightarrow t_A + \tau$ on
boundary A but which vanishes on boundary B.  By Noether's theorem, there
is a conserved generator $H_A$ of this symmetry which we may call
the Hamiltonian at boundary A.   Since the transformation vanishes
at boundary B and since bulk diffeomorphisms are pure gauge, on shell this
Hamiltonian is just a boundary term at boundary A. This last
statement is manifest in any on-shell covariant phase space
formulation (see e.g. \cite{symp1,WZ} for discussions based on symplectic
structures or \cite{HIM2} for a discussion based on the Peierls
bracket). In particular, one sees from e.g. \cite{HIM2} that $H_A$
is the sum of an integral of the usual boundary stress tensor
\cite{HS,kraus} over the hypersurface in boundary A defined by $t_A
= constant$ and some additional terms constructed from
$L_A$ at the same time $t_A$.   Since it generates a symmetry, $H_A$
is independent of the choice of $t_A$.

For later use it is convenient to construct the Hamiltonian using an
ADM-like canonical formulation.  We write the action in
canonical form by performing the usual space+time decomposition in
the bulk (see e.g. \cite{Waldbook}) and introducing canonical
momenta $p_A,p_B$ for the ancilla.  If the spatial manifold $\Sigma$
has boundaries $\partial_A \Sigma$, $\partial_B \Sigma$ where it
intersects the A- and B-boundaries, the result must take the
schematic form

\begin{eqnarray}
\label{Cact}
S_{total} &=& \int_{\Sigma \times {\mathbb R}}  \left(\pi \dot{\phi} - N {\cal H} - N^i {\cal H}_i \right) \cr &-& \int_{\partial_A \Sigma \times {\mathbb R}}   \left(N {\cal E}_A + N^i {\cal P}_{Ai} \right)   + \int dt_A \left(p_A \dot{\alpha}_A - \Delta_A \right) \cr
&-& \int_{\partial_B \Sigma \times {\mathbb R}}   \left(N {\cal E}_B + N^i {\cal P}_{Bi} \right)   + \int dt_B \left(p_B \dot{\alpha}_B - \Delta_B \right).
\end{eqnarray}
Here $\phi, \pi$ denote the full set of bulk fields and momenta,
including metric degrees of freedom, and a sum over fields is
implied.  The usual lapse and shift are denoted $N, N^i$, and ${\cal
H}, {\cal H}_i$ are the usual (densitized) bulk constraints, with $i$ running over directions on $\Sigma$.  The
boundary terms ${\cal E}_A$, ${\cal E}_B$, $ {\cal P}_{Ai}$, $ {\cal
P}_{Bi}$ are the boundary terms which would arise for $L_A,L_B =0$.
They depend only on the boundary values of $\phi,
\pi$, their derivatives along $\partial_A \Sigma$, and perhaps
certain coupling functions on the A- and B-boundaries. The
terms $\Delta_A$, $\Delta_B$ encode contributions from $L_A,L_B$.  As a result, they depend
on the respective ancilla ($\alpha_A, p_A$ or $\alpha_B, p_B$) as
well as boundary values of $\phi, \pi$, their derivatives along
$\partial_A \Sigma$, and any coupling constants present in $L_A,L_B$.
As for the bulk fields, $p_A \dot{\alpha}_A$ and
$p_B \dot{\alpha}_B$ are canonical ancilla kinetic terms and  a sum over all ancilla fields is implied.

We now consider any observable ${\cal O}(t_A)$ built from the boundary values of $\phi, \pi$ and the ancilla $\alpha_A, p_A$ at boundary time $t_A$.  It follows by direct calculation from (\ref{Cact}) that
\begin{equation}
\label{Aevolve}
\frac{d {\cal O} }{d t_A} = \{ {\cal O}, H_A  \} + \frac{\partial
{\cal O} }{\partial t_A},
\end{equation}
where $\frac{\partial {\cal O} }{\partial t_A}$ evaluates any
explicit dependence of ${\cal O}$ on $t_A$  and the A-boundary
Hamiltonian is
\begin{equation}
\label{HAapp}
H_A =  \int_{\Sigma }  \left(N {\cal H} + N^i {\cal H}_i \right) +  \int_{\partial_A \Sigma}   \left(N {\cal E}_A + N^i {\cal P}_{Ai} \right)  +  \Delta_A .
\end{equation}
Here we have assumed that $\partial_A \Sigma$ coincides with a surface of
constant $t_A, t_B$ on the A- and B-boundaries.  In \eqref{Aevolve} the lapse and shift
are arbitrary in the bulk and vanish on boundary B.  On boundary A,
the lapse and shift are dictated by the boundary conditions which
may force them to depend on the ancilla $\alpha_A, p_A$.  On-shell,
we have ${\cal H} = {\cal H}_i =0$ and the Hamiltonian is a pure
boundary term.    When the action is independent of $t_B$, a similar
result holds for the Hamiltonian $H_B$ which generates time
translations along boundary B while leaving boundary A unaffected.

We now wish to consider the case where the action does depend on $t_A$.  We note that any such action may still be written in the form \eqref{Cact}, with the only difference being that all coupling constants in ${\cal E}_A$, $ {\cal P}_{Ai}$, $\Delta_A$, may now depend on $t_A$.  Direct calculation now implies
\begin{equation}
\label{Aevolve2}
\frac{d {\cal O} }{d t_A} = \{ {\cal O}, H_A (t_A) \} + \frac{\partial
{\cal O} }{\partial t_A},
\end{equation}
with $H_A(t_A)$ again given by  \eqref{HAapp} evaluated at
A-boundary time $t_A$.  As desired, we see that this notion of
time-evolution is generated on-shell by a (time-dependent) boundary
term constructed only from A-boundary observables and Alice's ancilla
$\alpha_A, p_A$.

Although equations (\ref{Aevolve}) and (\ref{Aevolve2}) follow by direct computation from the action (\ref{Cact}), the reader may yet have a technical concern about our use of Poisson brackets.  In particular, the reader may note that coupling Alice's ancilla to the AdS system will require the boundary values of the gravitational field to become dynamical (see appendix B for a simple example involving scalar fields). The reader may then wonder whether the symplectic structure remains finite in such cases. Indeed, many familiar choices of gravitational symplectic structure (such as the explicit form given in \cite{WZ}) would diverge in this context.  Recall, however, that the symplectic form is not uniquely defined by the methods of \cite{WZ}, and in particular is ambiguous up to additions of an exact form $dB$ to the pre-symplectic form $\Theta$.  As shown in \cite{free}, one may make use of this ambiguity to define a new symplectic structure which remains finite under the desired conditions.  The relevant exact form $dB$ is closely related to the so-called counter-terms associated with what is known as holographic renormalization of the AdS gravitational action (see e.g. \cite{HS,kraus,KSthermo}).

\section{Time-dependent Boundary conditions: An example}
\label{exB}

It is perhaps enlightening to study a simple example which illustrates the physics of time-dependent couplings between an external system and bulk fields in an asymptotically AdS spacetime.  For simplicity and familiarity, consider a conformally-coupled scalar field $\phi_1$ in a fixed AdS background (AdS${}_1$).    In fact, it will be convenient to take the external system to {\it also} be a conformally-coupled scalar field $\phi_2$ living in a {\it different} AdS background (AdS${}_2$).  This second system is to be regarded as merely an example of the sort of ancilla that Alice might keep in her laboratory.

Since the fields are conformally coupled, we can instead describe the dynamics using rescaled scalars $\tilde \phi_1, \tilde \phi_2$ which propagate on, say, the north and south hemispheres of the Einstein static universe with line element

\begin{equation}
\label{dsESU}
d\tilde s^2 = \tilde g_{ab} dx^a dx^b = - dt^2 + d\theta + \sin^2 \theta \ d \Omega^2_{d-2},
\end{equation}
where $d \Omega^2_{d-2}$ is the line element on the unit $d-2$ sphere and where $\tilde \phi_{1,2}$ are defined on the regions $\theta \in [0, \pi/2]$ and $\theta \in [0, - \pi/2]$ respectively.  It will be convenient to denote the restriction of $\tilde \phi_{1,2}$ to the equator ($\theta=0$) by $\alpha_{1,2}$ and the corresponding normal derivatives at $\theta =0$ by $\beta_{1,2}$.  We take each normal derivative to be defined using the {\it outward}-pointing normal from the respective half of the spacetime, so that configurations symmetric under $(1 \leftrightarrow 2)$ and $\theta \rightarrow - \theta$ have $\beta_1 = - \beta_2$.

In order for the initial value problem to be well-defined, appropriate boundary conditions must be imposed on $\alpha_1, \alpha_2, \beta_1, \beta_2$.  We will specify such boundary conditions by first choosing an action for the system.  Consider for example

\begin{equation}
\label{csESU}
S_0 = - \int_{\theta > 0} \sqrt{\tilde g} \left( \frac{1}{2} (\partial \tilde \phi_1)^2 - \xi_{d} \tilde \phi_1^2 \tilde R  \right) - \int_{\theta < 0} \sqrt{\tilde g} \left( \frac{1}{2} (\partial \tilde \phi_1)^2 - \xi_{d} \tilde \phi_1^2 \tilde R  \right)
,
\end{equation}
where $\tilde R$ is the Ricci scalar of $\tilde g_{ab}$ and $\xi_d$ is the appropriate conformal coupling constant for spacetime dimension $d$.  Varying the action (\ref{csESU}) yields
\begin{equation}
\delta S_0 = \int_{\theta > 0} \sqrt{\tilde g} \ EOM_1 \ \delta \tilde \phi_1 + \int_{\theta < 0} \sqrt{\tilde g} \ EOM_2 \ \delta \tilde \phi_2 - \int_{\theta = 0} \sqrt{\Omega} \left(  \beta_1 \delta \alpha_1 + \beta_2 \delta \alpha_2    \right),
\end{equation}
where $EOM_{1,2}$ denote the usual conformally-invariant wave operators acting on $\tilde \phi_{1,2}$ respectively.  Thus, this action has well-defined variational derivatives if we impose boundary conditions fixing both $\alpha_1$ and $\alpha_2$.  In this case our two systems are decoupled and each satisfies an appropriate Dirichlet-type boundary condition.  In particular, each scalar has its own well-defined covariant phase space in which the symplectic structure is given by the associated (conserved) Klein-Gordon inner product.  Thinking of the two systems together as defining a single covariant phase space, the total symplectic structure is the sum of the two Klein-Gordon products.  As usual, the time-evolution associated with the $t$ coordinate of (\ref{dsESU})
is generated by the Hamiltonian
\begin{equation}
H_0 = \int_{t = constant, \theta > 0} \sqrt{\tilde g} \left(  \frac{1}{2} (\partial \tilde \phi_1)^2 + \xi_{d} \tilde \phi_1 \tilde R  \right) + \int_{t = constant, \theta < 0} \sqrt{\tilde g} \left(  \frac{1}{2} (\partial \tilde \phi_2)^2 + \xi_{d} \tilde \phi_2 \tilde R  \right)
\end{equation}
Note that we may fix $\alpha_{1,2}$ to be any (perhaps spacetime-dependent) function on the $(d-1)$-dimensional Einstein static universe at $\theta = 0$.

We now wish to couple our two systems at the $\theta = 0$ boundary by adding an interaction term to $S_0$.  Consider for example the action
\begin{equation}
\label{csESU2}
S_1 = S_0 +  \int_{\theta = 0} \sqrt{\Omega} \ f(x) \beta_1 \beta_2,
\end{equation}
where $f(x)$ is a fixed (i.e., field-independent) coupling function on the surface $\theta =0$ and $\sqrt{\Omega}$ is the volume element associated with the line element $d \Omega^2_{d-2}$.
Varying this action yields
\begin{eqnarray}
\delta S_1 &=& \delta S_0 +  \int_{\theta = 0} \sqrt{\Omega} \ f(x) (\beta_2 \delta \beta_1 + \beta_1 \delta \beta_2) \cr
&=& \int_{\theta > 0} \sqrt{\tilde g} \  EOM_1 \ \delta \tilde \phi_1 + \int_{\theta < 0} \sqrt{\tilde g} \ EOM_2 \ \delta \tilde \phi_2 - \int_{\theta = 0} \sqrt{\Omega} \left(  \beta_1 \ \delta \left(\alpha_1 - f(x)\beta_2 \right) + \beta_2 \ \delta \left( \alpha_2  - f(x) \beta_1 \right)  \right).
\end{eqnarray}
Thus the action $S_1$ yields a well-defined variational principle under boundary conditions which fix $\alpha_1 - f(x) \beta_2$ and $\alpha_2 - f(x) \beta_1$.  It is in this sense that the two systems are now coupled.

This coupled system has a well-defined covariant phase space with a well-defined Hamiltonian.  The symplectic structure is again the sum of the two Klein-Gordon inner products.  Now, however, neither Klein-Gordon product is conserved on its own.  Instead, there is a Klein-Gordon flux out of the $\theta < 0$ region proportional to
$F_{\theta < 0} = \int_{\theta = 0} \sqrt{\Omega} (\delta_1 \alpha_1 \delta_2 \beta_1 - \delta_2 \alpha_1 \delta_1 \beta_1 )$, and there is a similar flux out of the $\theta > 0$ region determined by $\delta_{1,2} \alpha_2, \delta_{1,2} \beta_2$.  But our boundary condition allows us to write
\begin{equation}
F_{\theta < 0} = \int_{\theta = 0} \sqrt{\Omega} f(x) (\delta_1 \beta_2 \delta_2 \beta_1 - \delta_2 \beta_2 \delta_1 \beta_1 ) = - F_{\theta > 0}.
\end{equation}
As a result, the total symplectic structure is conserved.  A straightforward computation of the Hamiltonian from (\ref{csESU2}) yields
\begin{equation}
H_1 (t) = H_0 + \int_{\theta = 0, t = constant} \sqrt{\Omega} \ f(x) \beta_1 \beta_2.
\end{equation}
It is easy to check that the above boundary condition removes all boundary terms from variations of $H_1$, so that we have a well-defined generator of time-translations as desired.  Other time-dependent couplings between bulk fields and external systems can be analyzed in a similar fashion.

As a particular application of the above framework, consider the case where $f(x)$ has compact support, so that the systems do not interact before some time $t_1$.  If we also take the initial state of the $\phi_2$-system to be excited in the distant past, this provides Alice with a certain amount of information and energy, some fraction of which will be injected into the (perhaps initially unexcited) $\phi_1$-system via the above coupling at around time $t_1$.   We note that, at the quantum level, the failure of the Klein-Gordon norms for $\tilde \phi_{1,2}$ to be separately conserved translates into a failure of unitarity for each system alone.  The two systems exchange information via the coupling, and only the coupled system evolves unitarily.   In much the same way, by considering similar couplings to other external systems, Alice can arrange to inject spins, radiation, or other quantum information into the AdS system.  In particular, the arguments of \cite{free} show that defining the coupling of Alice's ancilla to AdS boundary observables by writing down an action and choosing the AdS boundary conditions so that this action provides a well-defined variational principle will in general ensure that the total symplectic flux will be conserved, even in the presence of time-dependent couplings.   The results of \cite{free} also show that the appropriate symplectic structure remains finite even when the boundary values of the gravitational field become dynamical.

\end{document}